\documentclass[%
 reprint,
showpacs,preprintnumbers,
 amsmath,amssymb,
 aps,
prc,
superscriptaddress]{revtex4-1}

\usepackage{graphicx}
\usepackage{dcolumn}
\usepackage{bm}
\usepackage[colorlinks=true, pdfstartview=FitV, linkcolor=blue, 
citecolor=blue,urlcolor=navyblue]{hyperref}

\begin{document}

\title{Consistent relativistic mean field models constrained by GW170817} 

\author{Odilon Louren\c{c}o}
\author{Mariana Dutra}
\author{C\'esar H. Lenzi}
\affiliation{\mbox{Departamento de F\'isica, Instituto Tecnol\'ogico de Aeron\'autica, DCTA, 
12228-900, S\~ao Jos\'e dos Campos, SP, Brazil}}

\author{C\'esar V. Flores} 
\author{D\'ebora P. Menezes}
\affiliation{Depto de F\'{\i}sica - CFM - Universidade Federal de Santa Catarina, 
Florian\'opolis - SC - CP. 476 - CEP 88.040 - 900 - Brazil}

\date{\today}

\begin{abstract}
We have obtained the Love number and corresponding tidal deformabilites ($\Lambda$) associated with 
the relativistic mean-field parametrizations shown to be consistent (CRMF) with the nuclear matter, 
pure neutron matter, symmetry energy and its derivatives [Dutra {\it et al}., Phys. Rev. C 90, 
055203 (2014)]. Our results show that CRMF models present very good agreement with the recent data 
from binary neutron star merger event GW170817. They also confirm the strong correlation 
between~$\Lambda_{1.4}$ and the radius of canonical stars ($R_{1.4}$). When a recently GW170817 
constraint on $\Lambda_{1.4}$ and the corresponding radius $R_{1.4}$ is used, the majority of the 
models tested are shown to satisfy it.
\end{abstract}


\maketitle

\section{Introduction}

The discovery of the first binary pulsar PSR1913+16 by Russel Hulse and Joseph
Taylor in 1974~\cite{achar1} with its very stable and precise pulse period, and the 
observation in 1978  that its orbit period was declining with time~\cite{achar2},
opened a clear possibility  for the detection of gravitational
waves (GW). A probable explanation for the change in the period was the
loss of energy by the binary system in the form of GW. The detection
of these waves was expected since then until in
2015 the first signal was clearly seen (GW150914) and shown to be produced by
two colliding black holes~\cite{achar3}.  Finally, in 2017, LIGO and
Virgo made the first detection of GW170817 produced by colliding neutron
stars~\cite{PRL119} and the event was observed also as light in the optical, UV, IR,
X-ray and $\gamma$-ray emissions~\cite{AJL848}, what was then
called a multi-messenger observation. 

When one of the neutron stars in a binary system gets close to its companion just before
merging, a mass quadrupole develops as a response
to the tidal field induced by the companion. This is known as tidal
deformability~\cite{Damour86, Flanagan87} and can be used to constrain neutron star
macroscopic properties \cite{Piekarewicz2018, Providencia2018}, which
in turn, are obtained from appropriate equations of state (EOS). A nice and
simple review on the basic ingredients necessary to construct an EOS
is given in \cite{Hadrons2018}.

If one searchs the literature for an equation of state, hundreds of
models are found. Not too long ago, 263 relativistic mean-field (RMF) 
parametrizations were analyzed in~\cite{PRC90-055203} and confronted
with different sets of constraints, all of them 
related to symmetric nuclear matter, pure neutron matter, symmetry energy and its 
derivatives. The different sets differ from one another in the choice of validity ranges of certain 
quantities and in the level of restriction.  Only a small number of parametrizations of 
these models (35) were shown to satisfy adequately the chosen
constraints. This fact reinforces the idea that the proliferation of
models and the production of new parameter sets with a limited range 
of application should not be encouraged.

In~\cite{PRC90-055203}, the relativistic models were divided into~7 families, namely, 
linear finite range models (Walecka-type models~\cite{walecka}, type~1), nonlinear 
$\sigma$ models (Boguta-Bodmer models~\cite{boguta}, type~2), nonlinear $\sigma$ and 
$\omega$ models with a self-quartic interaction in the $\omega$ field (type~3),
nonlinear $\sigma$ and $\omega$ terms and cross terms involving these fields (type~4), 
density dependent models~\cite{NPA656-331} with couplings adjusted to nuclear properties 
(type~5), nonlinear point coupling models~\cite{PRC46-1757} (type~6) and models with 
$\delta$ mesons (type~7). Thirty of the approved models are of type~4, two are of 
type~5, one of type~6, and two of type~7 being both density dependent.

Later on, 34 RMF models that were shown to satisfy several nuclear matter constraints 
in~\cite{PRC90-055203}, namely, the same previous 35 models excluding the point-coupling 
one as discussed in~\cite{PRC93-025806}, were confronted with astrophysical constraints 
\cite{PRC93-025806}. The more important of these constraints are the
neutron stars with maximum mass in the range 
of \mbox{$1.93\leqslant M/M_\odot\leqslant 2.05$}~\cite{nature467-2010,science340-2013}, 
but the direct Urca process and the sound velocity also give hints on
the star cooling mechanism and its internal matter distribution.  
From the 34 analyzed models with nucleonic matter included, only 15
can sustain massive stars and none if hyperons are 
included, a result that accounts for the famous hyperon puzzle. Once hyperons are included 
in the calculations, the situation becomes more complicated because the EOS must be soft 
at subsaturation densities and hard at higher densities to predict massive stars, but 
hyperons soften the EOS. A possibility that reconciles the measurements of massive stars
with canonical stars with small radii present in the same family (another recently 
imposed constraint), is either the inclusion of strange mesons or of a new degree of 
freedom (not necessarily  known) in the calculations \cite{Luiz2018}. 

The measurements and analyses of data from this specific
gravitational wave established limits both on the dimensionless tidal
deformability of the binary system $\tilde \Lambda $ and on the
tidal deformability of the canonical  star $\Lambda_{1.4}$ as
being $\leqslant 800$ for the low spin priors upper boundary~\cite{PRL119} and 
contributed to the exclusion of very stiff EOS that would give rise to values larger than 
800. A lower limit was estimated  as $\tilde \Lambda > 400$~\cite{lower}. Recently, LIGO 
and Virgo collaboration updated the $\Lambda_{1.4}$ values to be constrained to the range 
of $70\leqslant\Lambda_{1.4}\leqslant580$~\cite{PRL121}. Moreover, the chirp mass, which 
relates the masses of both NS in the binary system  was observed to be  ${\cal M}=1.188 
M_\odot$~\cite{PRL119}.
The above mentioned boundaries combined with the chirp mass can be used to calculate
the bounds on the tidal deformability of the individual neutron stars
in the binary system~\cite{constanca}.

Since the detection of GW170817, several studies were dedicated to
look for correlations and sensitivity of important nuclear bulk
properties, i.e., the symmetry energy, its slope, compressibility and values of the
tidal deformability for the canonical $1.4M_\odot$ and other
slightly less and slightly more massive stars. In \cite{1805.00219}
the authors analyzed 4 Skyrme-type models and 1 obtained from a
density functional theory; in \cite{constanca} 18 relativistic and 24
nonrelativistic models were analyzed; in \cite{tsang} many Skyrme
type models were investigated and in \cite{1809.07108} 67 RMF models
were considered.  In the 3 last works, different correlations were
found between, for instance, $\Lambda$ and $R$,
$\Lambda_{1.4}$ and $R_{1.4}$, $\Lambda_{1.4}$ and $M_{max}$ or
$\Lambda_{1.4}$ and $\tilde \Lambda$.

In~\cite{PRL121}, a parametrized EOS was built at high-densities and one Skyrme EOS at
low densities and was confronted with GW170817 tidal deformability
information to obtain NS radii. The suggested values for the two
neutron stars in the binary system lie in the range
\mbox{$R_1=10.8^{+2.0}_{-1.7}$~km} and \mbox{$R_2=10.7^{+2.1}_{-1.5}$~km}. If a further
    restriction is imposed to account for EOS that support massive
    stars, both radii are constrained to the range of $11.9 \pm 1.4$~km. All models
analyzed in \cite{PRC93-025806} with a maximum mass of $(1.97 \pm 0.04)M_\odot$ or larger 
bear radii within the proposed range. In
\cite{1809.07108}, the authors established the upper limit on the
canonical stars radii as $R_{1.4} \leqslant 12.9$~km. In \cite{PRC93-025806},
only 7 models are excluded by this constraint.

Also, in a recent paper \cite{bao-an}, the authors show that an
infinite number of combinations of EOS with large slopes and small
compressibilities or small slopes and large compressibilities can lead
to the same $\Lambda_{1.4}$ and $R_{1.4}$, pointing to the need of
more observables so that the density dependence of the symmetry energy
be completely determined.

In all above mentioned papers, the models used to test constraints and
to look for correlations were randomly chosen. However, in
the present work we follow a more direct line of work by choosing
models we have already tested previously based on exactly the same constraints, 
avoiding models either with flamboyant degrees of
freedom or that have been forcefully corrected with extra mixed meson interactions
and return to the more conventional 34 RMF parametrizations that were
shown to satisfy the nuclear matter constraints
in~\cite{PRC90-055203}  to confront them with tidal deformabilities
 inferred from GW170817. 

\section{Results and discussion}

In a binary neutron star system, the tidal deformability is the
measurement of the perturbation generated by  the quadrupole moment in
one star as a response to the external field created by its 
companion. From the mathematical point of view, the dimensionless
tidal deformability, in terms of the Love number $k_2$, is given by
\begin{equation}
\Lambda= \frac{2k_2}{3C^5},
\label{tidal}
\end{equation}
where $C=m/R$ is the compactness of the neutron star of mass~$m$. The Love number $k_2$ is 
calculated by the following expression,  
\begin{align}
&k_2 =\frac{8C^5}{5}(1-2C)^2[2+C(y_R-1)-y_R]\times
\label{k2}
\nonumber\\
&\times\Big\{2C [6-3y_R+3C(5y_R-8)]
\nonumber\\
&+4C^3[13-11y_R+C(3y_R-2) + 2C^2(1+y_R)]
\nonumber\\
&+3(1-2C^2)[2-y_R+2C(y_R-1)]{\rm ln}(1-2C)\Big\}^{-1}, 
\end{align}
where $y_R\equiv y(r)$ is found from the solution of
\begin{equation}
r \frac{dy(r)}{dr} + y(r)^2 + y(r) F(r) + r^2Q(r)=0,
\label{ydef}
\end{equation}
with
\begin{equation}
F(r) = \frac{r - 4\pi r^3[ \epsilon(r) - p(r) ]}{r - 2M(r)}
\end{equation}
and
\begin{align}
Q(r)&=\frac{4\pi r\left[5\epsilon(r) + 9p(r) + \frac{\epsilon(r)+p(r)}{\partial 
p(r)/\partial\epsilon(r)}-\frac{6}{4\pi r^2}\right]}{r - 2M(r)} 
\nonumber\\ 
&- 4\left[ \frac{M(r)+4\pi r^3 p(r)}{r^2(1-2M(r)/r)} \right]^2.
\end{align}
The set of Eqs.~(1)-(5) can also be found in Refs.~\cite{new,tanija1,tanija2,binnington1,damour1}, 
for instance, in which earlier new developments were performed by using a large number of EOS's used 
to calculate $k_2$ and $\Lambda$.

Actually, Eq.~(\ref{ydef}) must be solved together with the well known TOV 
equations~\cite{tov}, 
in which $\epsilon$ and $p$ are the energy density and pressure, respectively, given as 
input. In our case, these quantities are given by the CRMF parametrizations with protons, 
neutrons, electrons and muons with the charge neutrality and $\beta$-equilibrium 
conditions together with the Baym-Pethick-Sutherland (BPS) equation 
of state~\cite{bps} in the low density regime, namely, 
$0.1581\times 10^{-10}$~fm$^{-3}\leqslant\rho\leqslant 0.008 907$~fm$^{-3}$. The initial 
condition for Eq.~(\ref{ydef}) is $y(0) = 2$ (related to the Love number order) and $M(r)$ 
is the neutron star mass enclosed within the radius $r$. At the surface of the star, in 
which $r=R$, one has $M(R)=m$. For detailed discussions on such calculations, we address 
the reader to Refs.~\cite{tanija1,tanija2,bharat1,binnington1,damour1}, for instance.

 The compactness of one recently measured  isolated neutron star~\cite{comp} is equal 
 to $0.105 \pm 0.002$. Notice that the GW170817 constrains NS in a binary system and 
hence, more measurements are necessary before this value is used as a constraint. 
Nevertheless, in Table \ref{tab1}, we show the compactness for the cases of 
$m=m_{\mbox{\tiny max}}$ ($C_{\mbox{\tiny max}}$), $m=1.4M_\odot$ ($C_{1.4}$), and 3 more 
cases obtained from the limits of possible masses in the binary system. The mass-radius 
diagrams used to calculate these $C$ values for the CRMF parametrizations are found in 
Ref.~\cite{PRC93-025806}.
\begin{table}[htb!]
\scriptsize
\centering
\caption{Compactness, in units of $M_\odot$/km, related to the maximum neutron star mass 
($C_{\mbox{\tiny max}}$), canonical one ($C_{1.4}$), and for $m_1 = 1.37$, $1.48$ and 
$1.60$ solar masses, with their respective values of $m_2$ for the CRMF models analyzed.}
\begin{tabular}{l|c|c|c|c|c|c|c|c|}
\hline\hline
Model & $C_{\rm max}$  & $C_{\rm 1.4}$ & $C^{m_1}_{\rm 1.37} $ & $C^{m_2}_{\rm 1.36}$ &  
$C^{m_1}_{\rm 1.48} $ & $C^{m_2}_{\rm 1.26}$ & $C^{m_1}_{\rm 1.60} $ & $C^{m_2}_{\rm 1.17}$ \\
\hline
BKA20 & $0.170$  & $0.105$  & $0.103$ & $0.103$ & $0.113$ & $0.095$ & $0.123$ &	$0.088$\\
BKA22  & $0.170$  & $0.105$ & $0.103$ & $0.102$ & $0.112$ & $0.094$ & $0.122$ & $0.087$\\
BKA24 & $0.169$  & $0.104$ & $0.102$  & $0.101$ & $0.111$ & $0.093$ & $0.121$ & $0.086$\\
\hline
BSR8  & $0.171$  & $0.108$ & $0.105$  & $0.105$ & $0.115$ & $0.097$ & $0.125$ & $0.090$\\
BSR9  & $0.170$  & $0.108$ & $0.105$  & $0.105$ & $0.115$ & $0.097$ & $0.125$ & $0.090$\\
BSR10  & $0.170$  & $0.107$ & $0.104$ & $0.104$ & $0.113$ & $0.095$ & $0.123$ & $0.088$\\
BSR11 & $0.169$  & $0.105$ & $0.095$  & $0.095$ & $0.112$ & $0.094$ & $0.123$ & $0.087$\\
BSR12 & $0.170$  & $0.106$ & $0.103$  & $0.102$ & $0.112$ & $0.094$ & $0.122$ & $0.087$\\
BSR15 & $0.160$  & $0.112$ & $0.109$  & $0.108$ & $0.119$ & $0.099$ & $0.132$ & $0.092$\\
BSR16 & $0.159$  & $0.111$ & $0.109$  & $0.108$ & $0.119$ & $0.099$ & $0.132$ & $0.092$\\
BSR17 & $0.159$  & $0.111$ & $0.108$  & $0.108$ & $0.119$ & $0.098$ & $0.131$ & $0.091$\\
BSR18 & $0.158$  & $0.110$ & $0.107$  & $0.107$ & $0.118$ & $0.097$ & $0.130$ & $0.090$\\
BSR19 & $0.158$  & $0.109$ & $0.106$  & $0.106$ & $0.117$ & $0.096$ & $0.130$ & $0.089$\\
BSR20 & $0.157$  & $0.107$ & $0.104$  & $0.104$ & $0.115$ & $0.095$ & $0.128$ & $0.087$\\
\hline
FSU-III & $0.158$ & $0.111$ & $0.108$ & $0.108$ & $0.119$ & $0.098$ & $0.132$ & $0.090$\\
FSU-IV & $0.160$  & $0.114$ & $0.111$ & $0.111$ & $0.122$ & $0.101$ & $0.135$ & $0.094$\\
FSUGold & $0.159$ & $0.113$ & $0.110$ & $0.109$ & $0.121$ & $0.100$ & $0.134$ & $0.092$\\
FSUGold4 & $0.160$ & $0.114$ & $0.111$ & $0.111$ & $0.122$ & $0.101$ & $0.135$ & $0.094$\\
FSUZG03 & $0.170$  & $0.108$ & $0.105$ & $0.105$ & $0.115$ & $0.097$ & $0.125$ & $0.090$\\
FSUGZ06 & $0.159$  & $0.112$ & $0.108$ & $0.108$ & $0.119$ & $0.099$ & $0.132$ & $0.092$\\
G2$^*$ & $0.176$  & $0.111$ &  $0.108$ & $0.108$ & $0.118$ & $0.099$ & $0.129$ & $0.092$\\
IU-FSU & $0.152$  & $0.112$ &  $0.109$ & $0.109$ & $0.118$ & $0.100$ & $0.129$ & $0.093$\\
\hline
Z271s2 & $0.153$  & $0.111$ &  $0.108$ & $0.108$ & $0.120$ & $0.098$ & $0.136$ & $0.090$\\
Z271s3 & $0.153$  & $0.114$ &  $0.110$ & $0.110$ & $0.122$ & $0.100$ & $0.139$ & $0.092$\\
Z271s4 & $0.153$  & $0.115$ &  $0.112$ & $0.111$ & $0.124$ & $0.101$ & $0.141$ & $0.093$\\
Z271s5 & $0.153$  & $0.116$ &  $0.113$ & $0.113$ & $0.125$ & $0.103$ & $0.142$ & $0.095$\\
Z271s6 & $0.154$  &$0.117$  &  $0.114$ & $0.114$ & $0.126$ & $0.104$ & $0.143$ & $0.096$\\
Z271v4 & $0.149$  & $0.115$ &  $0.111$ & $0.111$ & $0.124$ & $0.100$ & $0.147$ & $0.092$\\
Z271v5 & $0.149$  & $0.115$ &  $0.112$ & $0.111$ & $0.125$ & $0.101$ & $0.149$ & $0.092$\\
Z271v6 & $0.150$  &$0.116$  &  $0.113$ & $0.112$ & $0.126$ & $0.102$ &$0.150$ & $0.093$\\
\hline
DD-F & $0.193$ & $0.117$  & $0.114$  & $0.114$ & $0.125$ & $0.105$ & $0.137$ & $0.097$\\
TW99 & $0.196$  & $0.114$  &  $0.111$  & $0.111$ & $0.121$ & $0.102$ & $0.132$ & $0.095$\\
DDH$\delta$ & $0.192$ & $0.111$ & $0.109$ & $0.109$ & $0.118$ & $0.100$ & $0.127$ & $0.094$\\
DD-ME$\delta$ & $0.191$  & $0.118$ &  $0.115$  & $0.115$ & $0.126$ & $0.105$ & $0.137$ & $0.097$\\
\hline\hline
\end{tabular}
\label{tab1}
\end{table}

Models belonging to the same families present very similar compactness both for the 
maximum mass star and for the canonical one. In the low limit mass case, both stars of the binary 
system present practically the same mass, $m_1\approx m_2\approx 1.37M_\odot$. This is the 
reason they present the same compactness. However, this is  no longer true in the other 
cases, when one of the star is always more compact than its companion.

\begin{figure}[!htb]
\centering
\includegraphics[scale=0.34]{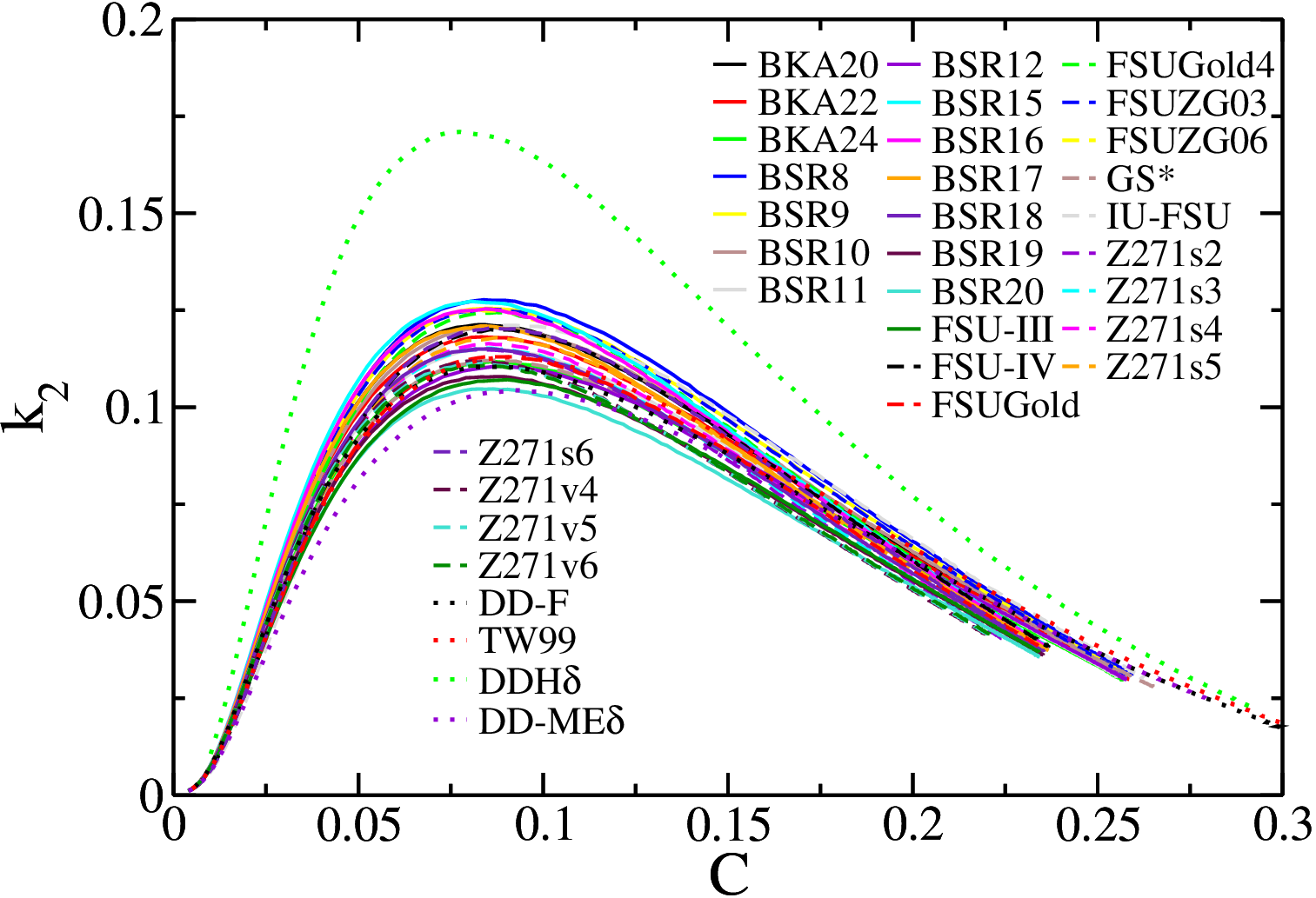}
\vspace{-0.2cm}
\caption{Love number $k_2$ as a function of the compactness for the CRMF parametrizations.} 
\label{k2fig}
\end{figure}

In Fig.~\ref{k2fig} we display the Love number $k_2$ for the CRMF parametrizations, obtained through 
the definition given in Eq.~(\ref{k2}) with $y_R$ calculated from the solution of Eq.~(\ref{ydef}) 
coupled to the TOV equations.
The pattern exhibited by $k_2$ is similar to that found in calculations involving other relativistic hadronic model, as 
one can verify in Ref.~\cite{bharat1}, for instance. It is important to point out that $k_2$ is very sensitive to the description of the crust of the star.
Once $k_2$ is calculated, it is possible to analyze the 
dimensionless tidal deformability by using the definition presented in  Eq.~(\ref{tidal}).

\begin{figure}[!htb]
\centering
\includegraphics[scale=0.34]{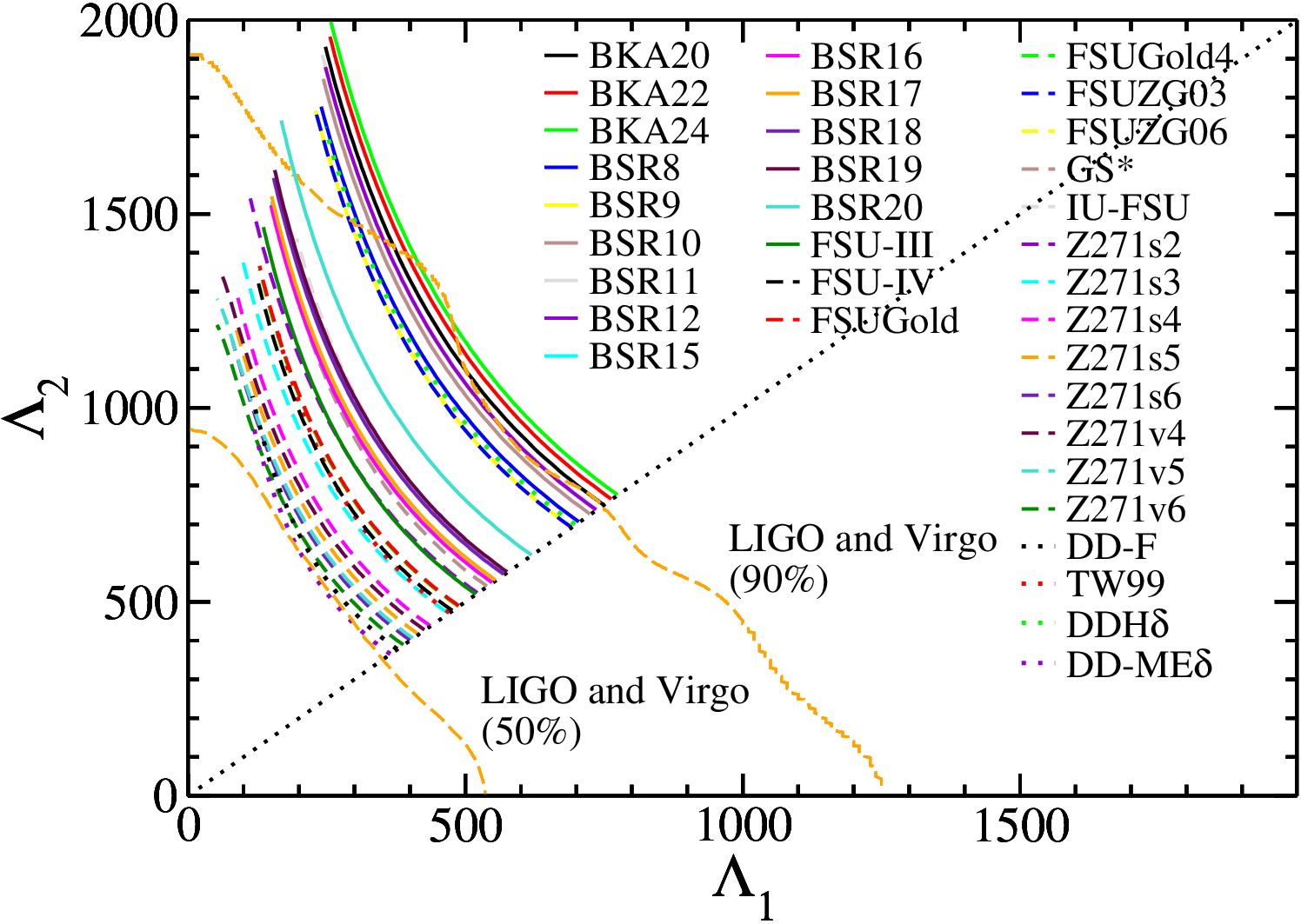}
\vspace{-0.2cm}
\caption{Tidal deformability parameters for both components of the observed GW170817. The 
confidence lines (90\% and 50\%) are the recent results of LIGO and Virgo collaboration 
taken from Ref.~\cite{PRL121}.} 
\label{def}
\end{figure}

In Fig.~\ref{def} we display a diagram of the dimensionless tidal deformabilities of 
each star in the binary system. $\Lambda_1$ is associated to the neutron star with mass 
$m_1$ 
which corresponds to the integration of every EOS in the range $1.37\leqslant 
m/M_\odot\leqslant 1.60$ obtained  from GW170817. On the other hand, the mass $m_2$ of the 
companion star is determined by solving the chirp mass $\mathcal{M} = \frac{(m_1 
m_2)^{3/5}}{(m_1 + m_2 )^{1/5}}$, whose value is $1.188M_\odot$, as determined in 
Ref.~\cite{PRL119}. 
One can see that all the investigated models lie in between the confidence
lines, what corroborates the fact that the previously constrained models to satisfy 
nuclear bulk properties are reliable to investigate neutron stars in binary systems, although many 
of them do not describe massive stars, as explained in the introduction of this letter. 

\begin{figure}[!htb]
\centering
\includegraphics[scale=0.34]{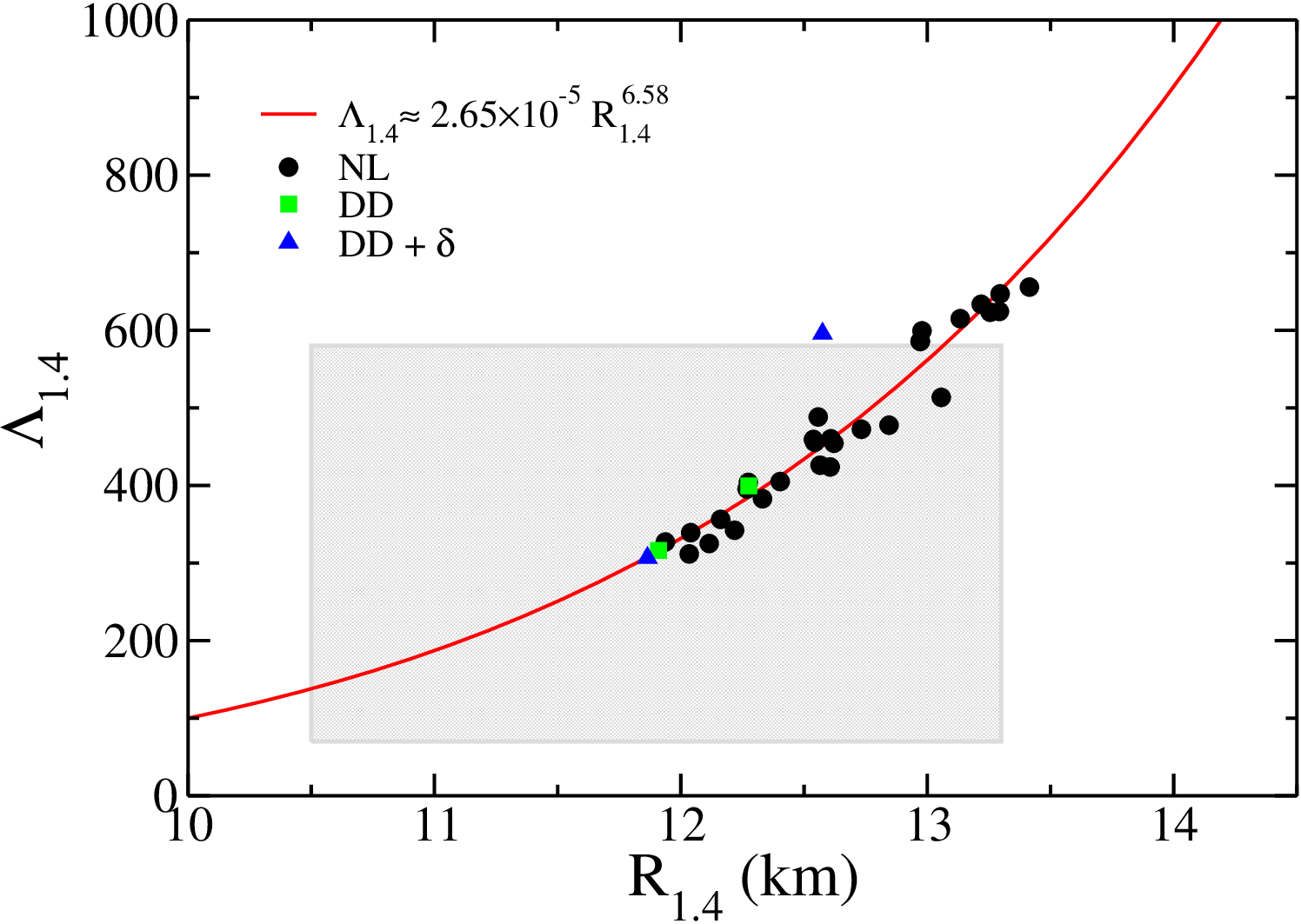}
\vspace{-0.2cm}
\caption{$\Lambda_{1.4}$ as a function of $R_{1.4}$, obtained from the CRMF models. NL, 
DD and \mbox{DD + $\delta$} stand for nonlinear, density dependent, and density dependent 
with $\delta$ particle, respectively. Gray area: results given in Ref.~\cite{PRL121}.} 
\label{def14}
\end{figure}

As already shown in Refs.~\cite{constanca,1809.07108,tsang}, we have also found a strong correlation 
between the tidal deformability of the canonical star and its radius both in linear and log scale 
(not shown), namely, $\Lambda_{1.4}=2.65\times10^{-5}R_{1.4}^{6.58}$, as can be seen in 
Fig.~\ref{def14}.
Since the second Love number $k_2$ depends on~$R$, for a given EOS, through a nontrivial differential equation coupled 
to the TOV one, $\Lambda_{1.4}$ as a function of $R_{1.4}$ is not simply given by $\Lambda_{1.4}\propto R_{1.4}^5$ as 
Eq.~(\ref{tidal}) suggests.

When we analyze the constraint for $\Lambda_{1.4}$ in the range 
\mbox{$70\leqslant\Lambda_{1.4}\leqslant 580$}, as proposed in~\cite{PRL121}, with 
corresponding $R_{1.4}$ values, depicted in Fig.~\ref{def14} by the shaded square, we 
observe that $24$ parametrizations (out of 34) are in accordance with this proposition. 
They are: BSR15, BSR16, BSR17, BSR18, BSR19, BSR20, \mbox{FSU-III}, \mbox{FSU-IV}, 
FSUGold, \mbox{FSUGold4}, FSUGZ06,  G2$^*$, \mbox{IU-FSU}, Z271s2, Z271s3, Z271s4, Z271s5, 
Z271s6, Z271v4, Z271v5, Z271v6, \mbox{DD-F}, TW99, and DD-ME$\delta$. It is interesting to 
notice that not all models capable of describing massive stars in the range 
\mbox{$1.93\leqslant M/M_\odot\leqslant 2.05$}~\cite{nature467-2010,science340-2013} 
discussed in \cite{PRC93-025806} lie inside the box with values obtained from GW170817, 
and not all the 24 models inside the gray area can describe massive stars. Only~5 models 
satisfy both constraints, namely, G2$^*$, \mbox{IU-FSU}, DD-F, TW99, and DD-ME$\delta$. 

Although the current range for $\Lambda_{1.4}$ is not very restrictive, we remind the reader that 
its values were still more imprecise, as one can verify in Ref.~\cite{PRL119} in which the range 
was computed as $\Lambda_{1.4}\leqslant 800$. Furthermore, there is a huge number of hadronic 
parameterizations coming from relativistic and non-relativistic models, around $500$ if we take into 
account only those from RMF and Skyrme models. Thus, it is important to find which particular set 
of parameterizations among these huge number is able to describe simultaneously different nuclear 
environments. In that sense, a constraint coming from the analysis of the recent GW170817, even being not 
so restrictive ($70\leqslant \Lambda_{1.4}\leqslant 580$) can be useful for this purpose.

\section{Final remarks}

In the present work, we have revisited 34 relativistic mean-field parametrizations 
shown to be consistent (CRMF) with the  nuclear matter, pure neutron matter, symmetry 
energy and its derivatives in \cite{PRC90-055203} and used them to compute the Love number 
and corresponding tidal deformabilities. We have checked that all analyzed models lie in 
between the confidence lines in the plot $\Lambda_2$ versus $\Lambda_1$. They also confirm 
previously obtained correlation between the tidal deformability and the radius of 
canonical stars. Once we use the GW170817 constraints on the tidal deformabilities to 
identify the corresponding neutron star radii range, as proposed in ~\cite{PRL121}, 24 
parametrizations are shown to satisfy them.  As far as the compactness, an important 
ingredient in the calculation of the Love numbers, is investigated, we have seen that, 
generally,  one of the star is always more compact than its companion, except in the low 
limit mass case $m_1=1.37$, when both stars in the binary system present the same 
compactness. 

It is also worth pointing out that only~5 parametrizations of the CRMF models, namely, 
G2$^*$, \mbox{IU-FSU}, \mbox{DD-F}, TW99, and \mbox{DD-ME$\delta$}, can simultaneously describe 
massive stars in the range \mbox{$1.93\leqslant M/M_\odot\leqslant 
2.05$}~\cite{nature467-2010,science340-2013}, as shown in~\cite{PRC93-025806}, and 
constraints from GW170817. 

To further constrain the existing EOS or confirm the results obtained so far, we look 
forward to the next detections of gravitational waves.

\section*{Acknowledgments} 
This work is a part of the project INCT-FNA Proc. No. 464898/2014-5,
was partially supported by CNPq (Brazil) under grants 301155.2017-8
(DPM) and 310242/2017-7 (OL), by Capes-PNPD program (CVF), and by Funda\c{c}\~ao de 
Amparo \`a Pesquisa do Estado de S\~ao Paulo (FAPESP) under the thematic project 
2013/26258-4 (OL,MD,CL).

\end{document}